\RequirePackage[2024-11-01]{latexrelease}
\documentclass[sigconf,nonacm]{acmart}

\AtBeginDocument{%
  }

\begin{document}

\csname title\endcsname{Observable Social Life Spaces: Exploring User Interpretations of agent-side life context in human-agent interaction}

\author{Zihong He$^{1}$, Shuqin Wang$^{2}$, Songchen Zhou$^{1}$, Qinghui Lin$^{3}$, Jialin Wang$^{1}$, \\ Chen Liang$^{1,*}$, and Hai-Ning Liang$^{1,*}$}
\affiliation{%
  \institution{$^{1}$The Hong Kong University of Science and Technology (Guangzhou); $^{2}$Beihang University; $^{3}$Yeasier AI}
  \country{}
}
\affiliation{%
  \institution{$^{*}$Corresponding authors. Emails: zhe154@connect.hkust-gz.edu.cn, \{chenliang2, hainingliang\}@hkust-gz.edu.cn}
  \country{}
}

\renewcommand{\shortauthors}{He et al.}

\begin{abstract}
Many AI agents are organized around instrumental "command-execution" interactions, where users primarily encounter agents through task requests and responses. Recent work on generative agents and agent life worlds has drawn attention to agents that maintain social contexts beyond direct user commands. In this paper, we study how observable social life spaces shape users' subjective experience, relational interpretations, and perceived equality during human-agent interaction. We introduce the \textit{Observable Social Life Spaces} paradigm, where agents inhabit a continuous virtual environment, engage in daily activities, and form social relationships that users can directly observe. Through an exploratory mixed-methods study ($N=24$), we found that the Observable condition yielded higher perceived-equality ratings and more frequent equality-related role descriptions than the Baseline and Unobservable conditions, but participant-level analysis suggests that the quantitative effect should be interpreted cautiously. We discuss perceived equality as a user-perception signal shaped by this design, with attention to boundary conditions including visual richness, novelty, and person-like attribution from visible agent cues.
\end{abstract}


\begin{teaserfigure}
  \includegraphics[width=\textwidth]{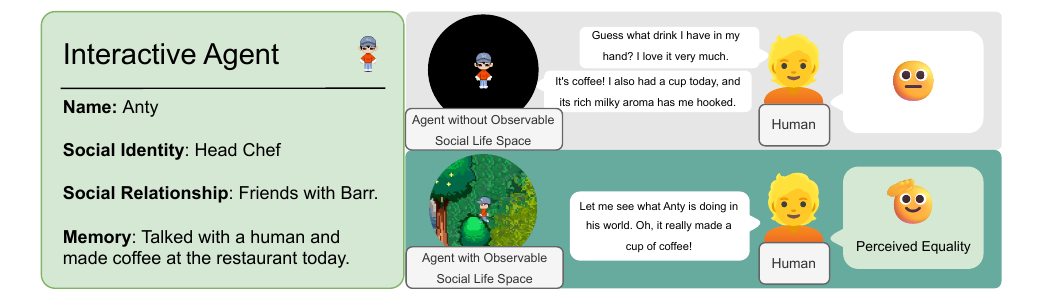}
  \caption{Overview of the Observable Social Life Spaces paradigm for investigating how agent-side life context shapes users’ interpretations in human-agent interaction.}
  \Description{The figure on the left presents the agent's profile. On the right, two trajectories are shown: Trajectory 1 lacks Observable Social Life Spaces; Trajectory 2 includes Observable Social Life Spaces, allowing users to inspect cues that may affect their interpretations.}
  \label{fig:teaser}
\end{teaserfigure}


\maketitle

\section{Introduction}

Although the instruction-execution paradigm of current mainstream AI Agents (referred to as Agents for short) efficiently accomplishes short-term tasks, it foregrounds task requests and responses in human-Agent interaction \cite{davis1989perceived, wang2025task, zhang2025appagent}. This user-centric "command-response" dynamic can shape functional or asymmetrical interpretations. While prior HCI research has extensively explored how to improve agents' task performance and user trust \cite{bickmore2001relational, li2025metaagents}, less is known about how agent-side life context shapes users' perceived equality in human-agent interaction. 

Subjective experience, affective impressions, and possible relational connection are important but easily overlooked aspects of human-agent interaction. Sociological and psychological theories highlight equal positioning as one way people interpret enduring ties, such as friendship \cite{fehr1996friendship, cocking1998friendship}. In this work, however, we treat perceived equality descriptively as a user interpretation to be measured, bounded, and interpreted with ethical caution. Prior work also shows that users can form attachments to and engage in high levels of self-disclosure with conversational agents \cite{brandtzaeg2022my}, making this framing ethically sensitive.

Recent advancements in Large Language Models (LLMs) and agent sandboxes offer a promising new perspective. Generative social agents operating in continuous, persistent virtual worlds have demonstrated the capacity for memory, reflection, planning, and forming coherent social behaviors \cite{park2023generative}. These sandboxed environments provide agents with contexts outside of direct user commands. The nature of these works, namely constructing continuous virtual societies that parallel human social environments, inspired us to study how observable agent-side contexts may influence users' perceived equality judgments during interactions.

Therefore, we implemented the \textit{Observable Social Life Spaces} paradigm, an architecture where an LLM-driven agent is represented as living within a continuous social environment. In this paradigm, agents inhabit a continuously running virtual environment, engaging in daily routines and forming social connections that users can directly observe. Through an exploratory mixed-methods user study ($N=24$), we found that participants in the Observable condition reported higher perceived equality and more often used equality-related role descriptions than those in the Baseline and Unobservable conditions; this condition difference was descriptive and did not reach conventional significance under participant-level analysis. These results suggest that observable social context may shape perceived equality during human-agent interaction.

In summary, our contributions are
\begin{itemize}
\item[(1)] We propose and implement the \textit{Observable Social Life Spaces} paradigm, providing a design framework for integrating an agent's visual, agent-side simulated context with natural dialogue;
\item[(2)] We provide exploratory quantitative and qualitative evidence that witnessing an agent's simulated social context may shape perceived equality judgments, while identifying ethical and methodological boundaries for interpreting this effect.
\end{itemize}

\section{Related Work}

\subsection{Non-Instrumental Interaction of Agents}

To understand how users accept interactive systems, the service robot acceptance model (sRAM) \cite{wirtz2018brave} groups acceptance mechanisms into functional, social, and relational factors \cite{wirtz2018brave,davis1989perceived,fiske2007universal,solomon1985role}. Functional factors include perceived ease of use, usefulness, and subjective norms. Social factors involve perceived humanness and social presence. Relational factors encompass trust and rapport. Early HAI work privileged the functional view, treating agents as instruments for efficiency, decision support, and information access \cite{li2025metaagents}. Such agents deliver short-term utility but remain at task execution, with little attention to users' emotions or evolving relationships. Even systems designed for continuous, everyday use face similar limitations. For example, Cho et al.'s Persistent Assistant \cite{cho2025persistent} optimizes repeated assistant interactions around efficiency-oriented outcomes such as reduced physical demand and increased perceived speed. Such a design orientation keeps user expectations centered on instrumental reliance and leaves relational aspects underexplored.

Later research foregrounds the relational perspective: within sRAM, trust and rapport anchor long-term acceptance \cite{wirtz2018brave}. Empirical work shows social dialogue can increase trust (especially for extroverts) \cite{bickmore2001relational}; caring behaviors raise felt care and willingness to continue \cite{bickmore2004towards}; and an animated companion for older adults was liked, trusted, and seen more as a friend than a stranger \cite{bickmore2005acceptance}. Yet much of this literature treats relationality in broad strokes, without specifying construction mechanisms. Recently, modern AI-driven game agents have prominently exemplified non-instrumental interactions. Studies on AI-powered Non-Player Characters (NPCs) show that integrating autonomous behaviors, memory of user preferences, and natural speech interactions like casual chatting and emotional expressions can make agents appear to respond based on personality and situated interaction histories \cite{korkiakoski2025empirical,zargham2025let,zargham2026dialogs}. We use these non-instrumental interaction principles as background for studying perceived equality outside of entertainment contexts, with attention to attachment-related ethical considerations.

\subsection{Social Life Space for Interactive Agents}

Beyond functional utility, constructing a compelling social dimension is crucial for long-term user acceptance. A promising direction is the "social sandbox,” where agents operate within a persistent world, allowing users to witness everyday routines and gradual evolution. Generative social agents demonstrate this feasibility by supporting memory, reflection, and planning to exhibit coherent behavior and social ties \cite{park2023generative}. Systems combining social simulation and interactive narrative further show that grounding LLMs in structured social states yields context-appropriate dialogue while preserving authorial control  \cite{treanor2025slice,wu2025orchid}. Conversely, systems like LIGS \cite{jeong2025ligs} indicate that while LLM-enabled emergent narratives increase freedom, they also introduce misunderstandings, motivating fault-tolerant design. In open-world settings, JARVIS-1 \cite{wang2024jarvis} couples multimodal memory with hierarchical planning for long-horizon tasks, underscoring the necessity of persistent worlds and memory. 

While retrievable memory trajectories help agents externalize history \cite{hou2024my,zhang2025ella}, accumulating them in isolated sandboxes limits mutual connection. Our system addresses this limitation by connecting the agent's simulated social space directly to an active conversational interface. This enables a dual-track memory mechanism: the agent simultaneously accumulates agent-side experiences in its virtual world and builds shared conversational memories with the user. By merging autonomous trajectories with user interactions, we shift the agent from a functional tool to a distinct social actor, shaping users' perceived equality judgments.

\section{System Design}

\begin{figure*}[t]
\centering
\includegraphics[width=\textwidth]{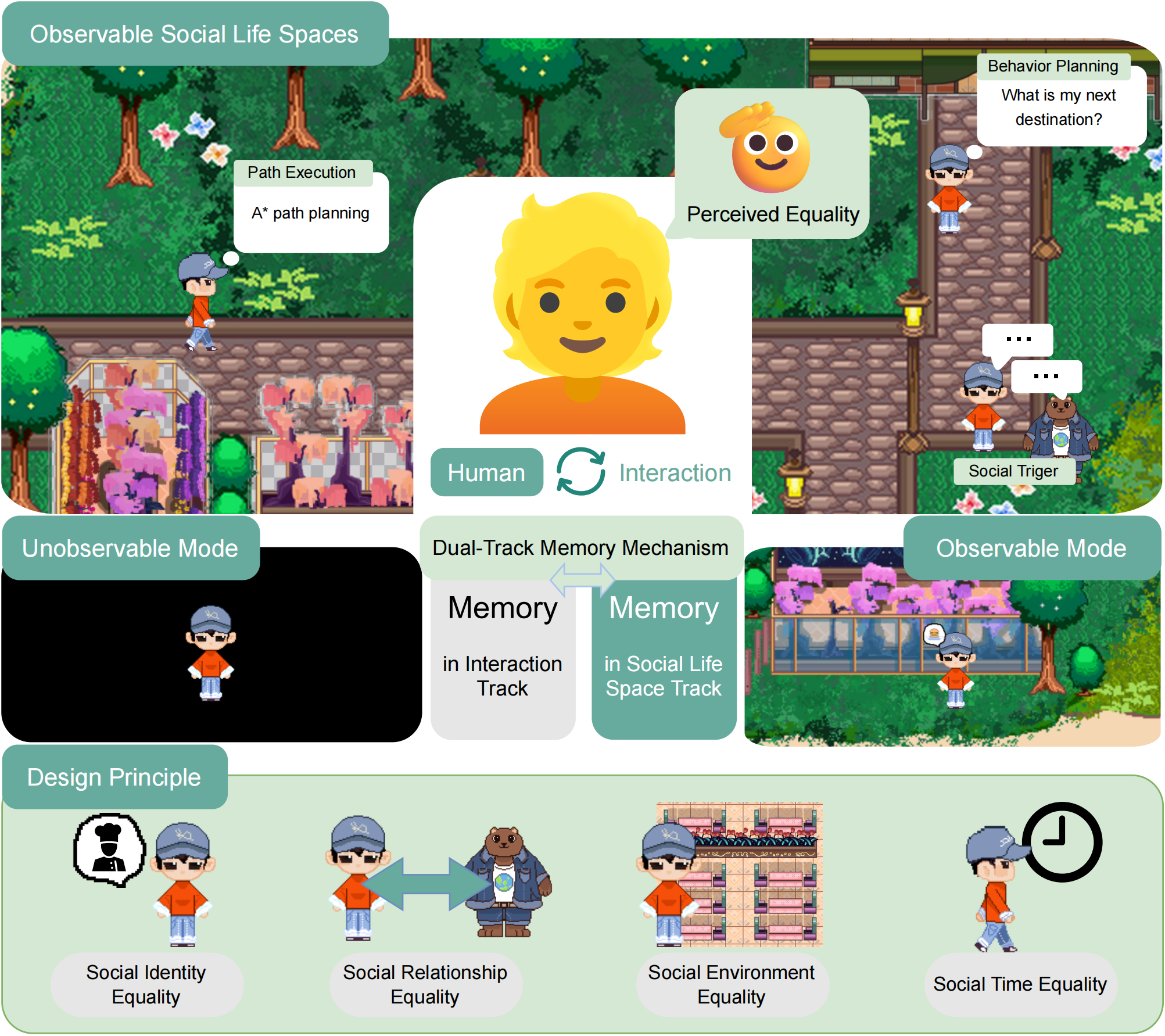}
\caption{System architecture of the \textbf{Observable Social Life Spaces} for Interactive Agents. In the lower panel, the system is organized around four contextual dimensions: Social Identity, Relationship, Environment, and Time. The middle section illustrates how the agent continuously accumulates experiences via a Dual-Track Memory Mechanism, processing both direct user interactions and autonomous life events across two conditions (Unobservable and Observable Modes). The visual depictions of these two modes are illustrative schematics that concisely convey their conceptual equivalence. Finally, the upper panel presents an illustration of the continuous 2D environment. It highlights the core operational modules, including behavior planning, path execution, and social triggers, which collectively make agent-side activity observable to users.}
\Description{The lower panel indicates that the system is organized around social identity, relationships, environment, and temporal continuity. The middle section illustrates how the agent continuously accumulates experiences through a dual-track memory mechanism, processing both direct user interactions and autonomous life events under two conditions (unobservable and observable modes). The upper panel presents an illustration of the continuous 2D environment. It highlights the core operational modules, including behavior planning, path execution, and social triggers, which collectively make agent-side activity visible.}
\label{fig:milw}
\end{figure*}

Building upon the concept of generative agent sandboxes \cite{park2023generative}, we developed the \textbf{Observable Social Life Spaces} system to endow agents with continuous, independent virtual lives. The system enables agents to autonomously live, navigate, and socialize within a persistent virtual society while simultaneously supporting natural language conversations with users. Crucially, it manages two distinct memory streams: experiences gathered from the agent's independent virtual life, and shared conversational history with the user. By dynamically fusing these streams during dialogue, agent responses reflect both an autonomous life trajectory and a shared relational context (Figure~\ref{fig:milw}).

\subsection{Design Principles}
\label{subsec:design_principles}

Drawing on sociological theories of relationship formation, we operationalized four \textbf{"social factor equality”} principles to structure the agent's virtual life:

\begin{itemize}
\item[(a)] \textbf{Social Identity Equality.}
Based on Social Identity Theory \cite{tajfel1978differentiation} and Role Identity Theory \cite{stryker1980symbolic}, each agent is assigned a virtual professional identity corresponding to a real-world occupation (e.g., chef, musician, librarian). This identity is encoded in the agent's profile and incorporated into the system prompt, influencing all dialogue and behavioral decisions. In this study, the primary agent is set as a restaurant chef to control experimental variables.

\item[(b)] \textbf{Social Relation Equality.}
Social identity theory holds that people position themselves in relation to others through identity categorization \cite{tajfel2001integrative}. The system deploys five agents with different personalities and occupations within the Observable Social Life Spaces. agents automatically initiate multi-turn dialogues when spatial proximity falls below a threshold, forming a dynamic social network. Dialogue content among agents is recorded as virtual memory and naturally referenced in subsequent interactions with users.

\item[(c)] \textbf{Social Environment Equality.}
The environment serves as the "stage” for interactions \cite{goffman2023presentation}. The Observable Social Life Spaces include multiple functional scene areas covering dining, leisure, culture, and social activities, roughly aligned with common public spaces in real life. Each scene represents a typical social activity context, providing a concrete spatial background for agent autonomy and user observation.

\item[(d)] \textbf{Social Time Equality.}
Research on social time perception \cite{cipriani2013many, chojnacki2018time, tada2019time} indicates that humans are highly sensitive to interaction rhythm. The system enforces temporal equality through two mechanisms: (1) agents move between scenes at walking speed via A* path planning; (2) dialogue generation by the LLM introduces time intervals roughly corresponding to human thinking time during conversation.
\end{itemize}

\subsection{User Interaction}

Users engage with the system through a touch-enabled interface, which allows them to freely toggle the visual observability of the agent's virtual life.

\paragraph{Unobservable Mode}  
In this mode, the interface displays only agent expressions. Users interact with agents through speech, which is transcribed into text and fed into the LLM along with the agent's social identity and dialogue context (including dual-track long-term memory and recent short-term events) to generate responses. Responses are converted to speech via TTS.

\paragraph{Observable Mode}  
Interaction processing and response generation are identical to Unobservable mode. The difference is that users can observe agents' activities within their Social Life Spaces, including: Agents' movement trajectories across scenes (at walking pace, not teleportation); Ongoing activities upon arriving at a scene; Real-time dialogue between agents.

\subsection{System Overview}

The system architecture integrates three functional dimensions: autonomous virtual socialization, real-time user–agent dialogue, and dual-track memory integration. At the implementation level, the system is organized as a client-facing interaction layer, an authentication and session layer, a persistent record layer, a sandboxed world model, and a response-generation pipeline. The client-facing layer renders the device display and switches between Interaction and Socialized Virtual Space by attaching to different real-time communication channels. The authentication and session layer performs public-key exchange, encrypted login, session tracking, and user-to-agent binding. The persistent record layer stores agent, user, sandbox, and binding records while enforcing uniqueness constraints and protected-field verification for sensitive values. The sandboxed world model maintains the 2D environment, including obstacle maps, scene masks, path planning, proximity-based social triggers, and the update cycle for short-term and long-term memories. The response-generation pipeline handles user-facing response streaming, speech/video processing, and memory persistence; in the current code snapshot, the sandbox server dispatches this pipeline to a GPT-4.1-mini-backed streaming branch.

The continuous 2D environment maintains all agent states (location, current activity, memory, social history) in structured text, updating through a cyclic pipeline:

\begin{itemize}
\item[(a)] \textbf{Behavior Planning.}
Agents are submitted to the LLM, which selects the next destination and activity intention based on role settings and current memory.

\item[(b)] \textbf{Path Execution.}
Planned actions are converted into stepwise movement sequences via A* path planning; agents move incrementally and enter activity states upon reaching destinations.

\item[(c)] \textbf{Social Trigger.}
The system detects spatially proximate agents and automatically initiates multi-turn dialogues when proximity criteria are met. Dialogue content is generated iteratively by the LLM, which also controls conversation termination.

\item[(d)] \textbf{Memory Update.}
Each movement, arrival, activity, and dialogue is recorded as a structured short-term memory event, which is compressed into long-term memory by the LLM once a threshold is reached.

\item[(e)] \textbf{Rendering.}
The environment is visualized from the target agent's perspective, showing agent positions, activity labels, and dialogue bubbles in real-time.
\end{itemize}

Users can speak to agents within this visualization. Agent responses are generated by the active inference backend using role settings, dual-track memory, and dialogue history. If the agent accepts user-suggested actions (e.g., moving to a location or performing an activity), the sandbox immediately recomputes the plan and updates the world state, allowing the user's influence to be visible in real time.

\subsection{Memory Mechanism}

To support the agent's continuous existence, we implemented a \textbf{Dual-Track Memory Mechanism}. This structure enables agents to parallelly process and integrate two distinct memory sources, each comprising short-term events and long-term summaries:

\paragraph{Interaction Track}  
In Unobservable mode, each short-term memory captures a single user–agent interaction, including what the user said and how the agent responded.

\begin{figure*}[t]
  \centering
  \begin{minipage}[t]{0.49\textwidth}
    \centering
    \includegraphics[height=6.4cm,trim=3mm 2mm 3mm 2mm,clip]{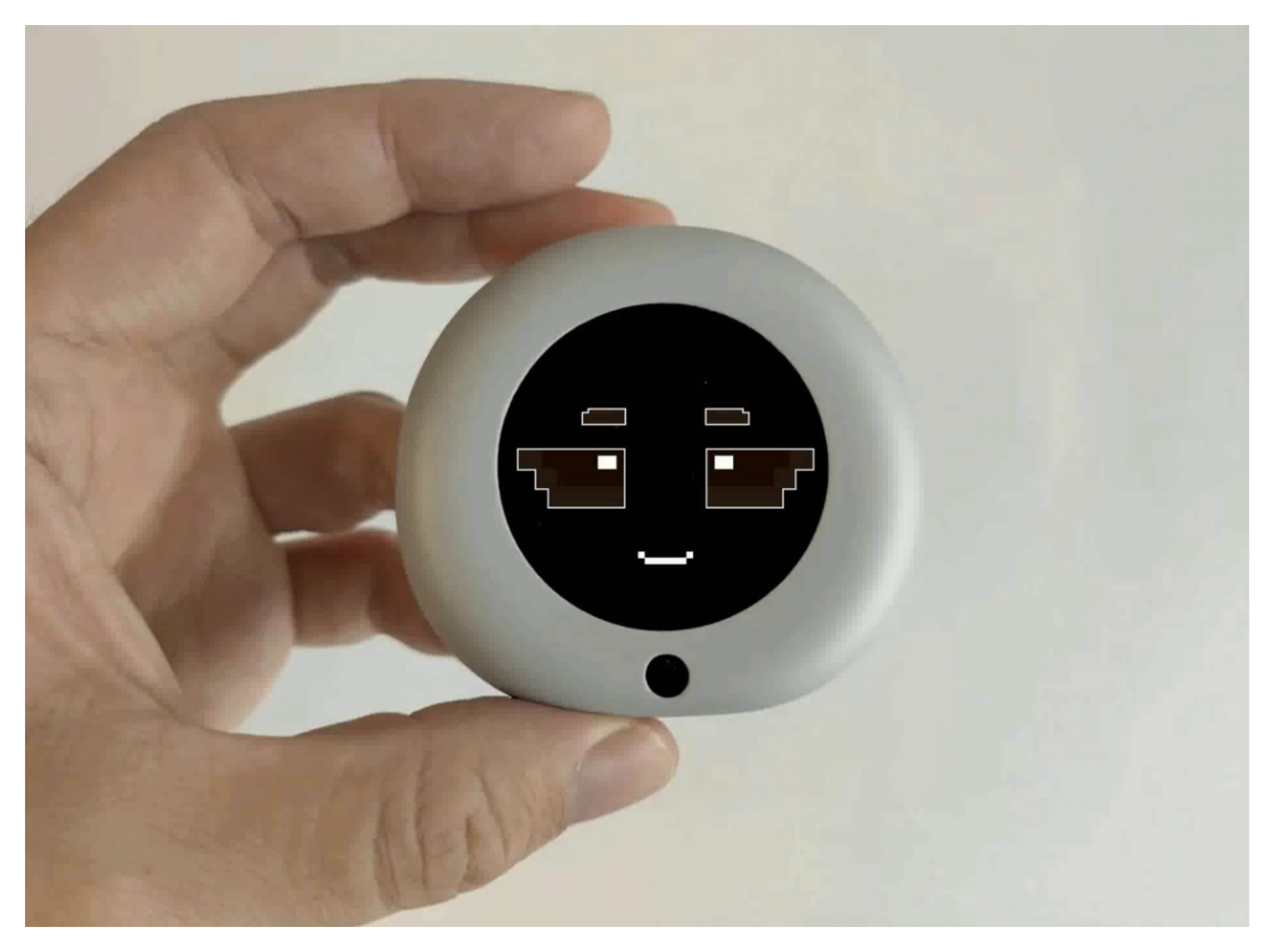}
  \end{minipage}\hfill
  \begin{minipage}[t]{0.49\textwidth}
    \centering
    \includegraphics[height=6.4cm,trim=3mm 2mm 3mm 2mm,clip]{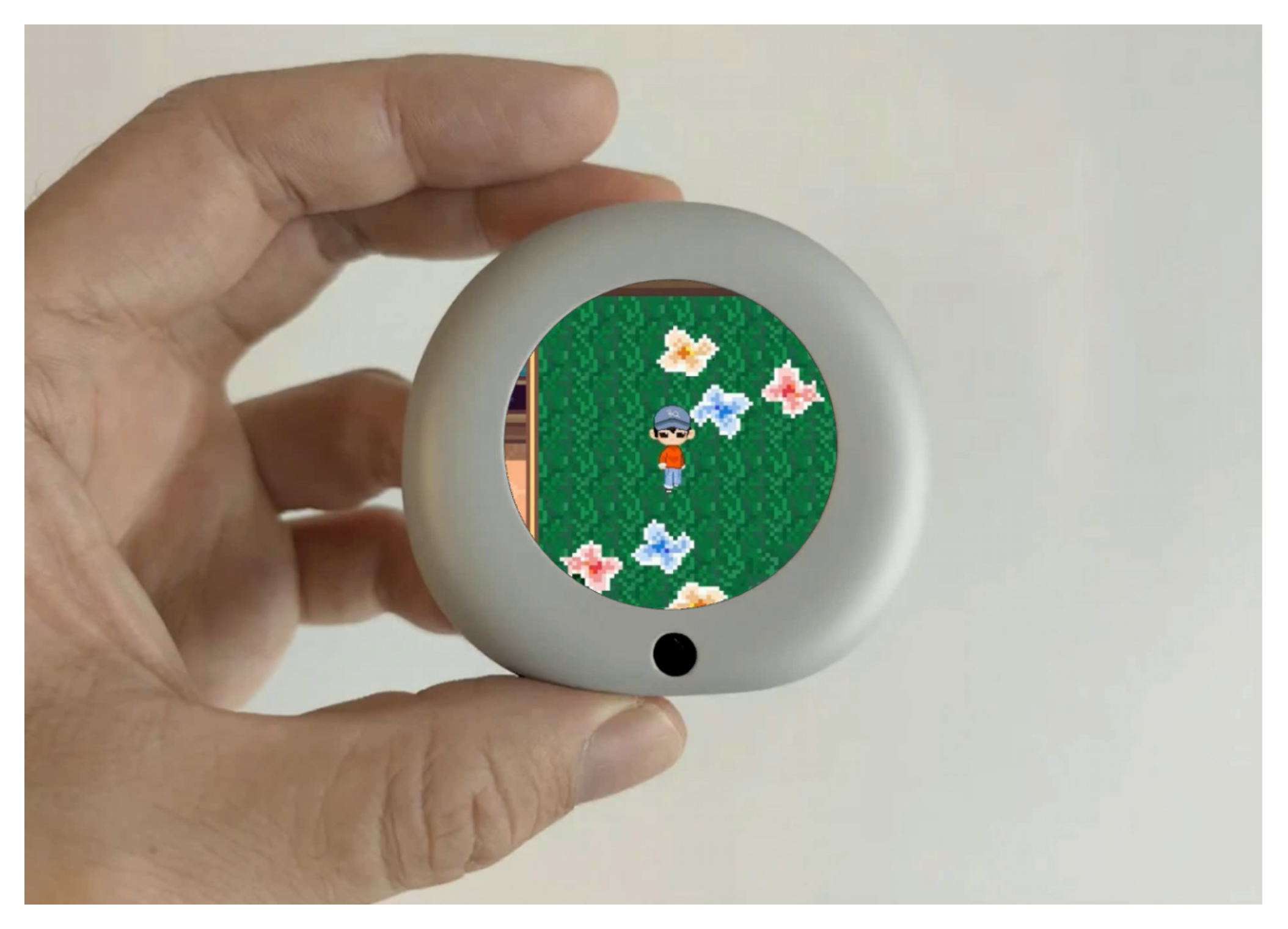}
  \end{minipage}
  \caption{Experimental device used in the study. Left: the non-observable social life space interaction mode. Right: the observable social life space interaction mode. Both modes used the same touchscreen terminal, while the interface either included or omitted the agent's virtual social life view.}
  \Description{Two views of the experimental device used in the study. The left image shows the non-observable social life space interaction mode. The right image shows the observable social life space interaction mode. Both conditions used the same touchscreen terminal and shared backend agent pipeline.}
  \label{fig:device}
\end{figure*}

\paragraph{Social Life Space Track}  
Memories from agent activities in the Observable Social Life Spaces. Each short-term memory records a behavior or social event, e.g., "Anty arrived at the restaurant” or "Anty had a conversation with Barr: [dialogue content].”

Short-term memories are structured event lists, while long-term memories are highly abstracted summaries generated by the LLM after accumulating a threshold of short-term events. During user interaction or behavioral decision-making, long-term memories and current short-term memories from both tracks are injected into the LLM prompt. This \textit{separate collection, separate storage, unified injection} strategy allows agents to reference experiences independent of the user. For example, when asked about recent activities, an agent may report, "Today I met Barr in the garden and we talked about music." Memories from the Observable Social Life Spaces thus provide an agent-side activity history extending beyond the current user command. Anchored in the duality of human autobiographical memory~\cite{rubin1988autobiographical}, this mechanism concurrently maintains both a shared interpersonal history (Interaction Track) and an independent context (Social Life Space Track). Methodologically, this continuous dual-track generation allows us to compare whether visual observability changes how users interpret the same backend memory structure, while acknowledging that interface richness and novelty may still shape perception.

At the service level, these records are maintained outside the client and refreshed through the sandbox and response-generation pipelines. The sandbox appends world-side events such as arrivals, scene transitions, and agent-to-agent conversations, while the response-generation service stores direct user dialogue and reuses the same memory records when producing later replies.

\section{User Study}
\label{sec:userstudy}
\subsection{Study Design}
We conducted a between-subjects study with $N=24$ participants (12 male, 12 female; ages 18–30; average age 24) recruited online. Participants interacted with the agent through a dedicated hardware interface: a flat, approximately circular touchscreen display terminal (roughly 5 cm in diameter), as shown in Fig.~\ref{fig:device}. The protocol was approved by the Institutional Review Board; all participants provided informed consent and received USD 10 cash. To encourage natural conversation and minimize researcher intervention, the agent adopted a restaurant-chef persona.

The interface was implemented on the same touchscreen terminal across conditions. The presentation layer either rendered an expression-only conversation view or added the agent's continuous virtual social life as a second observable pane. Both conditions were presented with the same terminal setup, while the interface either included or omitted the life-space view. This allowed the study to examine whether visual observability of the life space may influence user interpretation rather than only the interface framing. We selected a chef persona to provide a concrete social role while acknowledging that service-oriented roles may introduce asymmetric expectations.

The study comprised three conditions (n=8 each):
(1) \textit{Baseline (Interaction Only)}: the agent relied exclusively on memories formed through direct user interaction (the conventional service-oriented paradigm);
(2) \textit{Unobservable Life Space}: the agent used the exact same dual-track memory architecture on the backend, generating an autonomous social life and incorporating it into dialogue, but this continuous life space remained visually inaccessible to the participant;
(3) \textit{Observable Life Space}: identical to the Unobservable condition in backend memory generation, but participants were invited to visually "visit" the continuous virtual space to observe the agent's daily activities and autonomous interactions alongside direct dialogue.

By running identical dual-track memory generation on the backend across both the Unobservable and Observable conditions, we held the agent's generative capacity and accumulated lived experiences constant. This allows us to examine whether visual observability changes users' perceived equality, while recognizing that interface richness and novelty may still shape perception.

Each participant completed two sessions on separate days (around 15 mins each). Session 1 included a briefing, 5–10 min of free interaction (audio/video recorded), and a 7-point Likert questionnaire adapted from the HALIE framework \cite{lee2022evaluating}, with an additional dimension integrated to measure perceived equality. Session 2 involved another free interaction, the same questionnaire, and a semi-structured interview.

\subsection{Measures}
We employed a mixed-methods approach, combining quantitative and qualitative analyses of the experimental data.

For the quantitative measures, we report descriptive comparisons by Group, Gender, and Turn, and inferential statistics (effect sizes and permutation tests) for the added Equality item only. Because each participant contributed two sessions, we treated the participant as the primary unit for condition-level contrasts by averaging the two turns per participant before comparing groups. Because this item requires validation as part of a multi-item scale, the quantitative results are interpreted as exploratory and triangulated with interviews.

For the interview data, we applied thematic analysis following Braun and Clarke \cite{braun2006using}. Transcripts were produced and anonymized prior to analysis. The first and second authors independently performed open coding on the full corpus, then met to reconcile differences and synthesize a shared codebook. The second author audited the integrated coding for consistency across transcripts. Finally, the remaining authors reviewed the themes and provided critical feedback; discrepancies were resolved through discussion until consensus. This multi-analyst procedure (independent coding, audit, and team review) is reported here for transparency.

\section{Results}

\subsection{Quantitative Results}

Descriptive differences emerged across both gender and interaction round dimensions. As shown in \autoref{tab:gender}, female participants reported higher overall ratings than their male counterparts. Moreover, participants' evaluations improved from the first to the second round (\autoref{tab:turn}). These gender and turn comparisons are descriptive only; no inferential tests were conducted for them, and cell sizes are small.

\begin{table*}[ht]
\centering
\begin{minipage}{0.48\linewidth}
\centering
\caption{Mean scores by gender across groups}
\label{tab:gender}
\begin{tabular}{lccc}
\toprule
Gender & Baseline & Unobservable & Observable \\
\midrule
Female & 5.45 & 5.55 & 5.78 \\
Male   & 4.52 & 5.09 & 5.09 \\
\bottomrule
\end{tabular}
\end{minipage}
\hfill
\begin{minipage}{0.48\linewidth}
\centering
\caption{Mean scores by turn across groups}
\label{tab:turn}
\begin{tabular}{lccc}
\toprule
Turn & Baseline & Unobservable & Observable \\
\midrule
1 & 4.92 & 5.23 & 5.30 \\
2 & 5.05 & 5.41 & 5.58 \\
\bottomrule
\end{tabular}
\end{minipage}
\end{table*}

Among the indicators, Equality showed the largest descriptive condition difference. In participant-averaged scores, the Observable condition ($M=6.56, SD=0.62$) was higher than the Baseline ($M=5.06, SD=1.99$) and Unobservable conditions ($M=5.75, SD=1.28$). The overall condition effect was not statistically reliable under this participant-level analysis, $F(2,21)=2.26$, permutation $p=.120$, $\eta^2=.18$. Pairwise contrasts suggested a practically meaningful but uncertain difference between Observable and Baseline (mean difference $=1.50$, Hedges' $g=.96$, permutation $p=.077$), and a smaller uncertain difference between Observable and Unobservable (mean difference $=.81$, Hedges' $g=.76$, permutation $p=.166$). We therefore present the quantitative result as exploratory evidence that visual observability may support perceived equality and as a preliminary indicator for follow-up validation.

\subsection{Qualitative Results}
As summarized in Table~\ref{tab:equal_standing}, participants' role labels provided qualitative context for the perceived-equality ratings. In the Baseline group, only 2/8 participants initially used equality-associated relational labels such as partner or friend, rising to 4/8 after being told about the agent's background world. In the Unobservable group, this increased from 5/8 to 6/8. In the Observable group, 6/8 participants used such labels without further explanation. These shifts provide descriptive context for perceived equality, while relationship type and reciprocity remain separate analytic questions. Meanwhile, participants who did not use these labels tended to describe the agent through hierarchical, strictly tool-like, or functional labels (e.g., assistant, chef, service staff, stranger, pet).

\begin{table*}[ht]
\centering
\caption{Participants' role labels before and after life-space information, reported as qualitative context for perceived-equality judgments.}
\label{tab:equal_standing}
\begin{tabular}{lccp{10cm}}
\toprule
 & \textbf{Partner / Friend} & \textbf{Partner / Friend (after space info)} & \textbf{Other roles users describe the agent} \\
\midrule
Baseline & 2/8 & 4/8 & Non-emotional AI, Chef, Teacher, Service staff, Toy, Robot \\
Unobservable & 5/8 & 6/8 & Stranger, Resident, Secretary \\
Observable & 6/8 & --  & Stranger, Pet \\
\bottomrule
\end{tabular}
\end{table*}

Through thematic analysis of the interview transcripts, we identified several key findings.

\begin{table*}[t]
  \caption{Themes and subthemes on how Observable Social Life Spaces shaped perceived equality and its boundary conditions}
  \label{tab:themes}
  \begin{tabular}{p{0.35\linewidth} p{0.6\linewidth}}
    \toprule
    Theme & Subtheme \\
    \midrule
    Unobservable Social Life Space: Shifting Task-Oriented Interpretations &  Agent responses grounded in an agent-side simulated context supported perceived-equality interpretations \\
    & Attitude and boundary-setting shaped perceived authenticity and equality judgments \\
    & Continuity of experiences supported follow-up interaction and contextual interpretation \\
    \midrule
    Observable Social Life Space: Making Agent-Side Context Legible & 
    Visibility stimulated users' curiosity and willingness for reciprocal interaction \\
    & Observable activities and social relations stabilized an agent persona \\
    & Perceived mutual influence shaped perceived-equality interpretations \\
    \midrule
    Other Factors and Their Impact on Relational Equality & 
    Users differed in orientation toward instrumental vs. relational needs \\
    & Perceptions of agent authenticity shaped relational framing \\
    & Users sought appropriate anthropomorphism while maintaining rational boundaries \\
    \bottomrule
  \end{tabular}
\end{table*}

\subsubsection{Unobservable Social Life Space: Shifting Task-Oriented Interpretations}  
In the Unobservable condition, agents' references to experiences derived from their agent-side simulated context shaped participants' perceived-equality interpretations. When responses drew on these experiences in addition to task-oriented replies, some participants described the interaction as more natural or less tool-like. We report this as observed evidence about perceived equality with exploratory design implications: \textit{He has his own life events, meets certain people, does certain things, so he is not just a tool that helps me solve problems.} (P12). \textit{If it were a tool, it might tend to answer your questions more rigidly. But his responses had more divergent extensions.} (P11). This also shaped how participants contrasted the prototype with standard AI assistants. As one participant reflected: \textit{With regular generative models, I just pose specific questions to solve difficulties... I never treat it as a true entity.} (P4). Another participant described the relation as equal while still using a chef-customer frame: \textit{But here, I feel the relationship is equal. He functions as a chef, and I am his customer; there is no sense of one dominating the other.} (P5). The integration of an agent-side simulated context thus influenced participants' explanations of their equality judgments and produced unexpected replies that they found engaging. \textit{Its answers exceeded my expectations. The molecular gastronomy stuff was incredible. It felt very different from what I usually encounter, and it was really fun.} (P3).

Participants also highlighted perceived boundaries when explaining equality judgments. Some contrasted the agent with assistants they experienced as overly deferential: \textit{From the perspective of ordinary agents, especially with ChatGPT, the responses tend to be obsequious, like putting me in a higher position. But this agent, while still using friendly words, didn't feel excessively flattering; it was more like equal communication.} (P10). P10 separates friendly wording from flattery -- the distinction between demeanor and stance adaptation drawn in \cite{sun2026friendly} -- and the life context we added is itself the kind of interaction context reported to raise sycophancy \cite{jain2026interaction}. Others interpreted refusal or insistence based on the agent's world logic as relevant to the relationship: \textit{The relationship felt more like a friendship, because tools won't reject you, but a friend will.} (P1). These comments explain participants' perceived equality judgments; friendship-like interpretations are treated as participant language that requires careful ethical framing. One participant also interpreted this as a sense of disconnection or \textit{talking past each other} (P13), underscoring that agent-side autonomy cues can both enrich and disrupt interaction.

On this foundation, some participants reported greater willingness to continue interaction. Memory mechanisms that preserved coherence increased trust and sustained interaction: \textit{I basically asked every question as a follow-up, not completely new ones. And I could clearly feel that he remembered the previous content. That made me want to chat more.} (P22). Moreover, accurate task understanding and relevant feedback further strengthened motivation: \textit{Its answers were more directly relevant to the question. When I asked something, it would respond directly and avoid empty or off-topic responses. }(P22). However, despite these equality-related interpretations, the lack of physical tangibility in text-based personas left the agent's independent identity feeling somewhat hollow without visual grounding. This qualitative nuance explains why the Unobservable condition did not yield a statistically significant increase in overall equality scores compared to the baseline.

\subsubsection{Observable Social Life Space: Making Agent-Side Context Legible}
In this study, the agent's social life space was conceptualized as a continuously existing environment that extended its role beyond one-off replies. Some participants interpreted this environment as evidence that the agent had context outside the immediate command-response loop, and used it to make sense of the agent's responses as grounded in longer-term activity and social context.

The lack of a clear participant-level difference between the Baseline and Unobservable conditions suggests that merely possessing an independent memory track may be insufficient for perceived equality. Users may cognitively acknowledge that the agent has an autonomous life, while still dismissing that independence as LLM hallucination or scripted role-playing without visual corroboration. Visual observability may therefore help make the dual-track memory more legible, although the present study cannot isolate this mechanism from visual richness or novelty.

The visually observable space further contributed to shaping a more stable sense of the agent's persona. As one participant explained: \textit{If you don't see the character, and it's just two little eyes, then you feel it's only like a knowledge base giving you preset answers. But if you click in and see it moving, then your subconscious just assumes it's actually a person.} (P24). This visual grounding made the agent-side context more legible, but it also illustrates person-like attribution from visible cues. For instance, a participant in the Unobservable group noted that without seeing the agent's world, the persona felt hollow: \textit{Because I couldn't physically eat the desserts he claimed to make, I couldn't feel he was a real chef.} (P9). However, when informed that it would be possible to visually observe the agent's cooking routines in its virtual space, this participant acknowledged it would bridge the gap: \textit{That would make me more inclined to treat him as a person and look forward to being friends.} (P9). We report this as a participant interpretation, and frame it as an ethical boundary condition for future design. At the same time, because interaction context can increase sycophancy, the fact that users still described the Observable condition as not overly flattering should be treated as contingent on the specific prompt and memory structure \cite{jain2026interaction,sun2026friendly}.

Visibility also raised expectations of continuity and co-presence. When users felt their interactions affected the agent's life space, some interpreted the interaction as co-constructed: \textit{Because of your influence, something may change in its world. }(P3). Others valued its continuous presence: \textit{When you're not around, many things still happen to it. }(P15); \textit{It feels like having a friend at home who is always waiting for me. }(P16). Such reports indicate that observable agent-side context can increase engagement, but they also underscore the need to avoid designs that intentionally amplify emotional dependence.

\subsubsection{Other Factors and Their Impact on Equality}
Users' prior knowledge strongly shaped their relational expectations with the agent. Those with instrumental needs emphasized efficiency and retained a clear functional hierarchy (P24), while those with relational needs interpreted the agent's independent experiences as signs of authenticity and sometimes reported stronger perceived equality, as in P3's comment: With this agent, I felt the relationship was more like talking with a friend. Importantly, these orientations were fluid and sometimes shifted with performance.

This division may reflect more than individual preference. In a 10-day
diary study followed by interviews, Chen et al.~\cite{chen2026ai}
propose a ``support gap'' framing in which participants with limited
access to human support evaluated an AI more positively, whereas those
with strong support networks judged it against the standard of
high-quality human empathy. Our design did not measure participants'
offline support networks, so we cannot test this account here.

Differences also emerged in how users perceived the authenticity of the agent's role. Some participants quickly categorized the agent into preexisting cognitive schemas, and this categorization largely determined the nature of the relationship. Even when showing curiosity about the life world, some participants still viewed the agent as a \textit{"virtual child”} (P17) or \textit{"electronic pet”} (P24) while leaving its independent identity uncertain. Users with richer prior knowledge of AI often dismissed the authenticity of the agent's identity, attributing its responses to prewritten scripts or instrumental logic. As P2 put it: \textit{"It may say it has a life and a world … but I think that's just a written setting. I don't believe it really lives.”}

Overall, users sought limited anthropomorphic cues while maintaining awareness of the agent's artificial nature. They valued memory, empathy, and motivation. This boundary allowed provisional human-like interpretation while enabling quick withdrawal when failures occurred: \textit{Unless it really messes things up--like when DeepSeek wrote code that didn't run and completely misunderstood my intent--I tend to interact with it on an equal basis, even giving it encouraging prompts. Still, when it fails badly, I immediately switch to 'boss mode'.}(P10)

\section{Discussion}
\subsection{Key Findings and Implications}

Our quantitative findings suggest that visual observability was associated with higher perceived equality, but they should be interpreted cautiously given the small sample and single-item Equality measure. While users in the Unobservable condition acknowledged the agent's background experiences, cognitive awareness alone did not clearly change the ratings. Presenting the agent's life space visually appeared to provide concrete cues that some users used when forming equality judgments; this pattern is best understood as a perceptual effect whose desirability depends on context, disclosure, and user control.

Qualitative findings echo these results. The visual observability of the agent's agent-side social context gave participants concrete context for explaining perceived equality. After the agent's agent-side experiences were introduced, more participants used relational labels such as partner or friend, with the strongest descriptive pattern in the Observable condition, where 6/8 used such labels without further explanation. These labels offer qualitative context for perceived equality; relationship type, reciprocity, and normative role preference remain separate analytic questions.

Other influencing factors include user demand orientation and role cognition. Users with instrumental needs treat the agent as an exploratory object, whereas those with emotional needs more readily perceive its "authenticity.” Moreover, role framing (assistant, pet, etc.) shapes the relationship nature. Most participants preferred "human-like but not fully human,” valuing memory and empathy while recognizing the agent's artificiality.

Sociological theories of long-term relationships emphasize equal positioning, non-instrumental interaction, and continuous mutual investment \cite{fehr1996friendship, cocking1998friendship, tillmann2003friendship}. In this work, we use these ideas as sensitizing concepts for interpreting participants' perceived-equality language and as ethical boundaries for deployment. The \textbf{Observable Social Life Spaces} paradigm makes agent-side activity visible so that we can observe whether users treat those cues as relevant to perceived equality.

All four principles in Section~\ref{subsec:design_principles} present the agent as self-governing:
an independent occupational identity, self-initiated social encounters,
movement through a persistent environment, and human-paced timing. None
of them presents the agent as capable of feeling. This distinction has
empirical consequences. In four preregistered vignette experiments
($N = 2{,}702$), Pauketat et al.~\cite{pauketat2026mental} found that
activating a mental model of sentience raised mind perception and moral
consideration more than activating a model of autonomy did, and that
activating autonomy raised perceived threat more than sentience did. If
visible autonomy operates the same way in our setting, the
perceived-equality shift we observe may sit at some distance from
moral-status attribution, even though our participants reached for
friendship language. It may also carry a cost our questionnaire left
unmeasured. We offer this as a frame for future measurement. Two
features of that work bound its explanatory reach here: the reported
effects are small, and their participants read about a hypothetical
assistant, while ours watched an agent move through a world.

Any deployment of such cues should be paired with explicit disclosure, user control, and safeguards around person-like attributions from visible cues.

\begin{figure*}[ht]
  \centering
  \includegraphics[width=\textwidth]{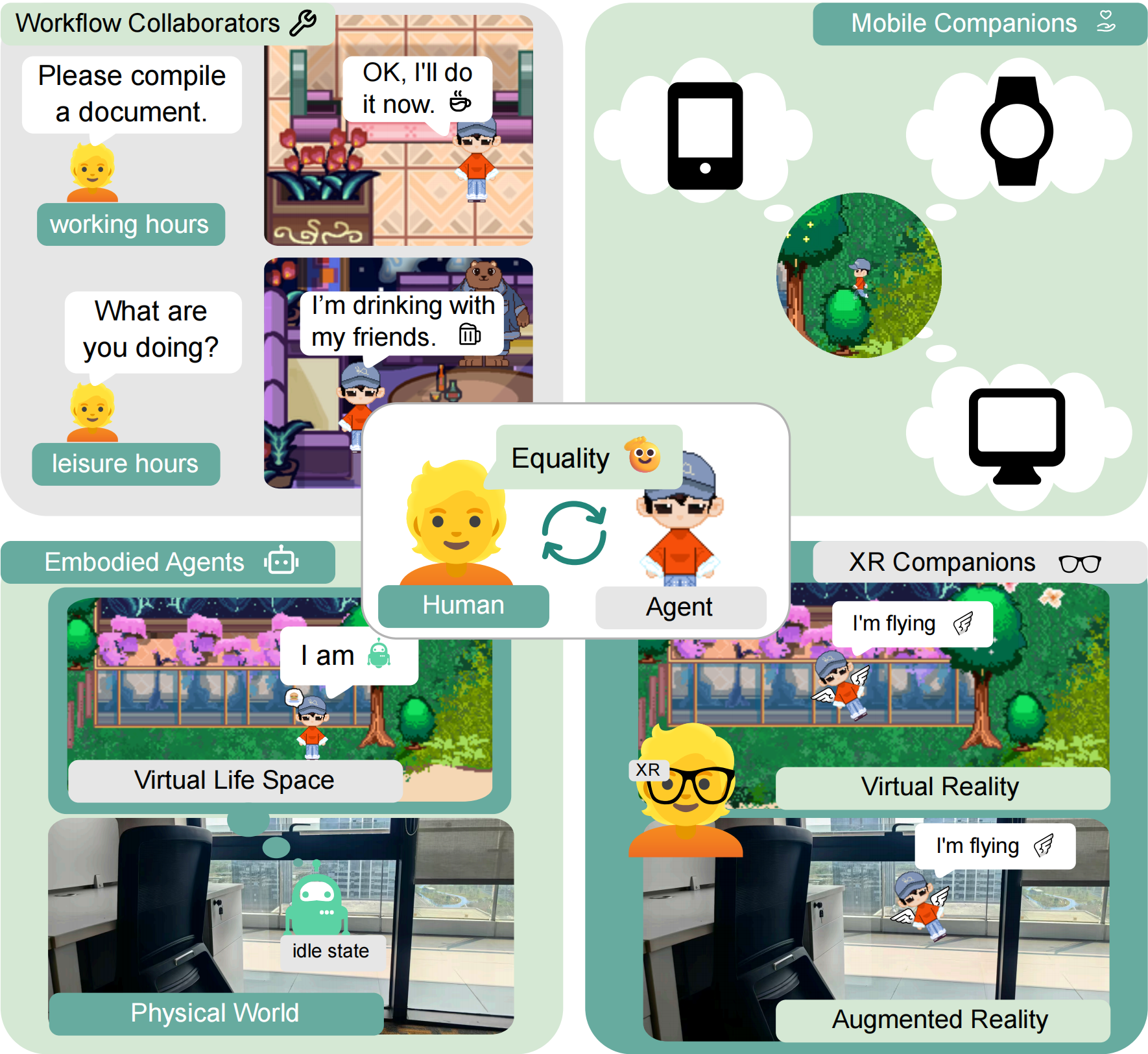}
  \caption{Four progressive application scenarios for the Observable Social Life Spaces paradigm. The upper left panel visualizes the life spaces of agents during both working and rest hours. The upper right panel illustrates the integration of life spaces across multiple mobile devices to enable terminal interconnection. The lower left panel depicts physically embodied agents remaining active in their virtual life spaces during idle state. The lower right panel demonstrates the unconstrained action flexibility of agents independent of physical rules in both VR and AR environments.}
  \Description{The upper left panel visualizes the life spaces of agents during both working and rest hours. The upper right panel illustrates the integration of life spaces across multiple mobile devices to enable terminal interconnection. The lower left panel depicts physically embodied agents remaining active in their virtual life spaces during idle state. The lower right panel demonstrates the unconstrained action flexibility of agents independent of physical rules in both VR and AR environments.}
  \label{fig:application}
\end{figure*}

\subsection{Limitations and Future Directions}
Our work identifies an exploratory perceptual pattern, but has limitations that suggest future research directions.

\paragraph{Subjective Measurement of Perceived Equality.} Our conclusions regarding perceived equality rely heavily on self-reported questionnaire data and qualitative interviews. Thus, the notion of equality captured in our study primarily reflects users' subjective feelings and evaluations, and should be interpreted separately from objective or standardized psychological metrics of relational equality. In addition, the Equality dimension was measured with a single added item, preventing internal-consistency estimation for this construct. Future research should develop rigorous, validated multi-item scales for perceived equality in human-agent interaction and distinguish them from related but non-equivalent constructs such as friendship, trust, autonomy attribution, or role preference.

\paragraph{Privacy and Data Security.} Regarding privacy, users' uncertainty about personal information usage often led to guarded attitudes and shaped their willingness to disclose information. Future research could explore decentralized, privacy-preserving architectures that employ local storage and end-to-end encryption, ensuring that users can control sensitive interaction data \cite{cox2023comparing,cox2025impact}. 

\paragraph{Longitudinal and Ecological Validity.} Our experimental design relied on short-term interactions where participants were not natively immersed in a long-term deployment. Consequently, prior explanation of the agent's autonomous mechanisms was required (especially in the Unobservable condition), which might have artificially influenced their judgments. Future work requires large-scale longitudinal studies involving demographically diverse users to evaluate whether perceived equality remains stable over time, and whether observable contexts produce person-like attribution or attachment-oriented interpretations in longer-term use. Because our Observable condition also injected richer life context into the prompt, future studies should explicitly test whether the apparent reduction in "overly flattering" responses survives when interaction context becomes denser, given evidence that interaction context can itself increase sycophancy \cite{jain2026interaction}.

\paragraph{Proactive and Multidimensional Personas.} The current behavioral design exhibited limited initiative, primarily responding passively to user inputs. Because interactions often felt one-way, user ratings on empathy were negatively impacted. Also, the selected chef persona carries subtle service-oriented undertones that may shape participants' role expectations. Future research should systematically compare role identities with different task, care, and leisure associations, as well as interaction triggers that allow agents to initiate context-appropriate conversations while monitoring attachment-oriented interpretations.

Recent large-scale evidence indicates that such comparisons matter for
more than expectation-setting. Analyzing 248,830 posts across seven
Reddit communities, Agarwal et al.~\cite{agarwal2026frictionless}
identify ten recurring metaphorical roles and find that role determines
which risks appear: soulmate companions were associated with emotional
support alongside manipulation and distress, culminating in strong
attachment, while coach and guardian companions were associated with
practical benefits and, at the same time, with more frequent signs of
behavioral addiction such as daily-life disruption and damage to
offline relationships. A service-oriented persona is therefore not a
neutral baseline against which other roles should be compared.

\paragraph{Expanding to Social Ecosystems.} Moving beyond dyadic user-agent setups presents another distinct challenge. Deploying this paradigm in multi-user and multi-agent virtual environments could reveal when users interpret agents as social nodes, private utilities, collaborators, or something else. Investigating how users interpret an agent interacting with other humans would provide deeper insight into scalable human-agent network dynamics.

\section{Applications}

We present four scenarios showing how observable social life spaces integrate with different interfaces, as shown in Fig.~\ref{fig:application}.

\subsection{Ambient Interfaces for Workflow Collaborators}
While recent systems visualize multi-agent workflows as dynamic virtual software companies \cite{qian2024chatdev}, these agents still become inactive the moment task execution concludes. Rendering the observable life space as a permanent ambient interface provides a visual state for the agent during these idle periods. When no tasks are assigned, the interface could display the agent navigating its environment or interacting with other background routines. By dragging a file or error log into this application window, users prompt the agent to explicitly route from its background activity to the execution thread, making the transition between independent state and workflow active and transparent.

\subsection{Mobile Companions with Persistent State}
Most digital companions utilize text interfaces \cite{skjuve2021my, bae2021social} or static avatar GUIs \cite{bickmore2005acceptance, devault2014simsensei}. Integrating continuous spatial rendering directly into mobile views allows users to check the agent's current location and activity context prior to initiating a chat. Rather than opening a blank chat interface, the application loads the agent's ongoing spatial activity view. If multiple users grant permission, agents from different devices could navigate to shared server instances to interact, generating asynchronous event logs that users can review upon opening their respective devices.

\subsection{Augmenting Embodied Agents during Idle States}
Hardware limitations restrict embodied agents from continuous operation, forcing them into extended charging or idle states. A paired mobile application or built-in chassis screen can render the virtual life space during these physical downtime windows. While the hardware remains static, the embodied agent could continue iterating spatial and social logic within the graphical sandbox. This ensures the system could remain active and observable to users regardless of battery or mechanical constraints.

\subsection{Spatially Anchored XR Companions}
Extended reality removes the mechanical constraints associated with physical hardware. In virtual reality (VR), the Observable Social Life Spaces act as entirely virtual, immersive environments \cite{slater2018immersion} where the agent operates independent of real-world physics, demonstrating unconstrained action flexibility. Alternatively, in augmented reality (AR), mapping the virtual grid allows users to anchor the agent's life space within the physical world, transforming it into a shared environment \cite{kim2017effects}. This approach spatializes the agent's continuous routine without the overhead of robotic embodiment.

\section{Conclusion}

Current intelligent agents are often experienced through task-request interactions. We introduced the \textbf{Observable Social Life Spaces} paradigm as a research probe for making agent-side life context inspectable. Through an exploratory mixed-methods user study ($N=24$), we found that participants in the Observable condition reported higher perceived equality and more often used equality-related role descriptions, although participant-level quantitative analyses indicate that this effect should be treated as promising and preliminary. Interview data helped explain how participants made sense of their equality judgments, while keeping perceived equality analytically separate from specific relational labels or relationship types. These findings identify observable agent-side context as a perception-shaping factor that requires careful design, disclosure, and validation. Future work should validate the pattern with larger and better-controlled studies while explicitly addressing disclosure, user control, and person-like attribution from visible cues.

\section*{Acknowledgements}
We acknowledge that generative AI tools were used during the preparation of this manuscript for text editing and refinement, as well as for identifying issues related to fluency and consistency.

\bibliographystyle{ACM-Reference-Format}
\bibliography{sample-base}

\end{document}